\documentclass[%
 reprint,
 amsmath,amssymb,
 aps,
]{revtex4-1}

\usepackage{graphicx}% Include figure files
\usepackage{dcolumn}% Align table columns on decimal point
\usepackage{bm}% bold math
\begin{document}

\preprint{APS/123-QED}

\title{Quantum interactions, Predictability and Emergence Of Gravity}% Force line breaks with \\

\author{Vyshnav Mohan}

 \email{vyshnavmohan@iisc.ac.in}
\affiliation{%
 Indian Institute of Science, Bangalore 560012, India \\
 }%

\begin{abstract}

In this paper, we will show that gravity can emerge from an effective field theory, obtained by tracing out the fermionic system from an interacting quantum field theory, when we impose the condition that the field equations must be Cauchy predictable. The source of the gravitational field can be identified with the quantum interactions that existed in the interacting QFT. This relation is very similar to the ER= EPR conjecture and strongly relies on the fact that emergence of a classical theory will be dependent on the underlying quantum processes and interactions. We consider two concrete example for reaching the result - one where initially there was no gravity and other where gravity was present. The latter case will result in first order corrections to Einstein's equations and immediately reproduces well-known results like effective event horizons and gravitational birefringence.

\end{abstract}

\pacs{Valid PACS appear here}% PACS, the Physics and Astronomy
                             % Classification Scheme.
%\keywords{Suggested keywords}%Use showkeys class option if keyword
                              %display desired
\maketitle

%\tableofcontents

\section{\label{sec:level1}Introduction}

Our understanding of gravity has undergone dramatic changes in the last couple of years by the advent of ideas like gravity/gauge theory correspondence and ER= EPR conjecture \cite{PROP:PROP201300020,VanRaamsdonk2010,PhysRevLett.96.181602, Maldacena1999}. However, we are still far away from exploiting these relations to its full potential. Much of the familiar examples of such correspondences come from string theory. In this paper, we will show how gravity can emerge out of an effective filed theory when we demand that the theory has to be Cauchy predictable. 

 Existence of a well-posed initial value problem goes to the heart of classical physics. Imposition of such a constraint on an effective filed theory will force that the underlying metric to be modified.  However, this condition will immediately put a constraint on the dynamics of the gravitational field as shown in Ref. \cite{fps14, Rat10, kris12}. This is because suitable initial data surface are required to uniquely evolve the initial data. We will explicitly calculate the modified Lagrangian for the gravitational dynamics in two particular situations.

It turns out that we can, in fact, identify the source of the emergent gravitational field with the quantum interactions that were traced out to get the effective field theory. This draws a parallel between the quantum correlations of a field theory with a classical theory of gravity. This is very similar to the entanglement and gravity equivalence that has already been conjectured by Susskind, Maldacena, Van Raamsdonk and others.

Another important aspect of the calculations carried out in the paper is that there is neither strings on the gravity side nor a gauge theory in the  't Hooft limit in the field theory side.

  In section II, we will consider Euler-Heisenberg action and show that corrections to the flat space metric emerges. In section III, we will consider QED on a curved background and explicitly calculate the modified Einstein's equations that we get from the theory and solve it for a spherically symmetric spacetime.
\section{ QED on a flat spacetime}
 Quantum electrodynamics on a flat spacetime has been studied extensively for a very long period of time. Let us begin by considering the Euler-Heisenberg action \cite{physics/0605038}. In the lower energy limit, we can vary the action to get the equation of motion given by:

 \begin{eqnarray}
 D_{a} F^{a b} -\dfrac{1}{m^4} [16 z F^{a b}F_{c d}D_{a} F^{c d}  -8 y (F^{a c}F_{c d}D_{a} F^{b d} +\nonumber \\
\hspace{1 cm}F^{a c}F^{b d}D_{a} F_{c d})]  \label{three} ~ ~~~~~.
\end{eqnarray} 
Here, $ z =\frac{-\alpha^2}{36}$ , $\zeta=-\frac{ 7\alpha^2}{96}$ where $\alpha$ is the fine structure constant and $m$ is the mass of the electron. All the physical quantities are in natural units.

It is easy to see that photon propagation in this theory will be different from that dictated by Maxwell's equation. In fact, we will specifically see how the tensor that raises and lowers the indices will get modified. For that purpose, we use geometric optics approximation. Here the electromagnetic field tensor is assumed to be a product of a slowly varying amplitude and a rapidly varying phase, given by \cite{sch92,shore03}
\begin{equation}
 F_{\mu \nu} =  f_{\mu \nu} e^{i \theta}
\end{equation}

 and the wave vector is given by $k_{\mu} = \partial_{\mu} \theta$. Together with Bianchi identities $D_{[\mu} F_{\lambda \nu]}=0 $, we get the following constraint : 
\begin{equation}
  f_{\mu \nu} =  k_{\mu}a_{\nu} -  k_{\nu}a_{\mu}
\end{equation}
 Where $ a^{\nu} $ is a vector in the direction of the polarization of the photon and can be assumed to obey the relation $k_{\mu}a^{\mu}= 0$. The polarization 4-vector is assumed to be space like normalized with respect to $\eta_{ab}$ and the latin indices run from $ 0$ to $3$. Thus Eq.~(\ref{three}) becomes :

\begin{equation}
  \eta_{a c}k^{a}k^{c} -\dfrac{8}{m^4} (4 z +y ) F_{a b}F_{c d}k^{a}k^{c} a^{b}a^{d}  -\dfrac{8y}{m^4} (F_{a}^{ d} F_{c d}k^{a}k^{c} ) = 0 \label{mod}
\end{equation} 

Thus, the modified light cone condition given by Eq.~(\ref{mod}) can be written as 
\begin{equation}
  G_{ac}  k^{a} k^{c} = 0
\end{equation}
\begin{equation}
{ \text{Where}  \hspace{.09cm}  G_{ac} := \eta_{a c} -\dfrac{8}{m^4} (4 z +y ) F_{a b}F_{c d} a^{b}a^{d}  -\dfrac{8y}{m^4} (F_{a}^{ d} F_{c d} ) } \nonumber
\end{equation}
This is strikingly similar to the light cone condition in general relativity, given by:
\begin{equation}
  g_{ac}  k^{a} k^{c} = 0
\end{equation}
 
 Now we turn our attention to something very interesting. It can be shown that if we are given a dispersion relation for some matter field and if we demand that these fields have to be Cauchy predictable, then the principle polynomial defined in [Ref.\cite{Rat10,fps14}] has to be a bi-hyperbolic and energy distinguishing homogeneous polynomial. From the light cone condition or from the coefficients of the highest-order derivative in Eq.~(\ref{mod}), we can read off the dual polynomial for the system as
\begin{equation}
 {  P^{\#} (x,v) = G_{ab} v^{a} v^{b} }.
\end{equation}
Certain properties of $G_{ab}$ follows directly from the definition of the tensor. We find that $G_{ab}$ is a billinear, symmetric, non- degenerate map on that tangent space $T_{x}M$ of the manifold $M$ at every point $x$. Thus, $G_{ab}$ by itself can be treated as an effective metric of the manifold, provided an inverse metric exist at all points on the spacetime. It turns out that it is even possible to define $G _{ab}$ as a metric tensor of the manifold and find appropriate $g_{ab}$ which solves for $G _{ab}$ and satisfies its  non-degeneracy condition. 

Thus, we calculate the principal polynomial to be
\begin{equation}
 {  P (x,k) = G^{ab} k_{a} k_{b} }.
\end{equation}
 where $ G^{ab} $ is the inverse of the metric $ G_{ab} $. Another compelling reason to consider $ G_{ab} $ as an effective metric is the following:

 The wave vector $k_{\mu}$ is defined as a one-form belonging to the cotangent space $T_x^{\ast}(M)$  at every point of spacetime while the momentum $p^{\nu}$ of the photon belongs to the tangent space $T_{x}(M)$. In this modified theory, the relationship between them is not trivial as  $ p^{\nu} = \eta^{\mu \nu} k_{\mu}$. Instead the wave vector is mapped to the momenta up to a factor by the Gauss map (See Ref.\cite{Rat10}) given by $[k] \mapsto \bigl[\dfrac{\partial P}{\partial k_{a}}(x,k)\bigl] $ where $[X]$ represents the projective equivalence class of all vectors collinear with X. Thus, we see that the correct relationship between $ p^{\nu}$ and  $k_{\mu}$ is:
\begin{equation}
p^{a} =  \dfrac{\partial ( G^{ab} k_{a} k_{b})}{\partial k_{a}} = G^{ab} k_{b}
\end{equation}
 This shows that $G^{ab}$ takes the role of $\eta^{ab}$ in raising and lowering the indices of vectors and co-vectors. Thus, $G^{ab}$ can be treated as an effective metric to the manifold.

For the modified Maxwell's equations to have a well posed initial value problem, the principal polynomial should be hyperbolic. Since the underlying geometry is determined by an effective metric $ G^{ab} $, the hyperbolicity of dual polynomial reduces to the condition that there exist some vector field $h$ such that $ P^{\#}(h)  >0$ and for any vector field q, the following equation has only two real roots
\begin{equation}
 {  P^{\#} (x,h + \lambda q) =0 }.
\end{equation}
This condition will be satisfied if $G^{ab}$ has a lorentzian signature.

With the kinematical structure of the theory determined, it is necessary to evolve the geometric initial data among hypersurfaces so that all them serve as initial data surfaces for the modified matter field equations. From the geometrodynamical point of view, this information is encoded in the dynamics of the gravitational field. The lagrangian of gravitational dynamics can be explicitly calculated from a set of linear homogeneous partial differential equations given in Ref.\cite{kris12,fps14}, which are obtained from the modified matter equations and it has, a priori, nothing to do with the geometric metric.  Thus, the imposition of Cauchy predictability of the modified matter equations leads to the modified gravitational dynamics.

 Obtaining the Lagrangian from the principal polynomial is quite tedious. However, as the kinematical structure of the spacetime is described by a metric $G_{ab}$, we can use the hindsight from Einstein-Hilbert action that the gravitational part of the new Lagrangian must contain a term proportional to the Ricci scalar of the metric $G_{ab}$. It is also possible to reach the same conclusion by traversing the steps given in Ref.\cite{fps14}. Thus, we arrive at the new lagrangian. 
\begin{equation}
 \therefore \mathcal{L}= \int  \left[ {\dfrac{1}{2\kappa}} \left( \mathcal{R} - 2 \Lambda \right) + \mathcal{L}_\mathrm{M} \right] \sqrt{-g}\, \mathrm{d}^4 x   \hspace{.09cm},  \text{where}  \hspace{.09cm}{\large{\kappa = 8 \pi G}}
\end{equation}
$\mathcal{R}$ is the Ricci tensor obtained from the metric $ G_{ab}$, $\Lambda$ is the cosmological constant and $ \mathcal{L}_\mathrm{M}$ is the Lagrangian of the matter fields appearing in the theory. We will vary the action explicitly for another system in the next section.

 This is a very interesting and surprising result. We have started off with Electromagnetic field on a flat spacetime and the imposition of Cauchy predictability of the effective one-loop theory has resulted in a curved spacetime! We will discuss more about this in section IV and see how it is related to the quantum interactions.
\section{ QED on curved spacetime}

The next example that we will be focusing on will be the Drummond and Hathrell effective action $\Gamma_{\text{DH}}$, which can be calculated by accounting for one-loop corrections in QED \cite{drum80, shore03}  i.e.
\begin{equation}
\Gamma_{\text{DH}} = \Gamma_{M} +\text{ln det S}(x, x^{'})                                                                                                                                                                                                                                                                                                                                                                                                                                                                                                                                                                                                                                                                                                                                                                                                                                                                                                                                                                                                                                                                                                                                                                                                                                                                                                                                      
\end{equation}

Here, $\Gamma_{M}$ is the free Maxwell action and S$(x, x^{'})$ is the electron propagator on the curved spacetime. Suppressing higher order derivatives of the curvature tensors relative to $O({\lambda}/{L})$, where $\lambda$ is the photon wavelength and $L$ is the typical curvature scale, we can find out the effective action as in Ref.\cite{drum80}.
After varying the action, we obtain the following modified Maxwell's equation \cite{shore03}:
\begin{equation}
D_{\mu} F^{\mu \nu} -\dfrac{1}{m^2} [2\sigma R_{\mu \nu}D^{\mu} F^{\lambda \nu} + 4 \zeta g^{ \nu \kappa} R_{\mu \kappa \lambda \rho}D^{\mu} F^{\lambda \rho}]  \label{eq:three}
\end{equation}
Here, $ \sigma =\frac{13\alpha}{360\pi}$ , $\zeta=-\frac{ \alpha}{360\pi}$ where $\alpha$ is the fine structure constant and $m$ is the mass of the electron. All the physical quantities are in natural units and $ F^{\mu \nu}$ is the electromagnetic field tensor as usual.

As in the previous section, we use geometric optics to get the modified light cone condition:
\begin{equation}
 { g_{ab} k^{a} k^{b} +  {\dfrac{2 \sigma }{m^2}} \hspace{.09cm}   R_{ab} k^{a} k^{b}-  {\dfrac{8 \zeta}{m^2}} \hspace{.09cm}   R_{acbd}k^{a} k^{b}a^{c} a^{d} =0 }.
\end{equation}
Here, the polarization 4-vector is assumed to be space like normalized with respect to $g_{ab}$ and the latin indices run from $ 0$ to $3$. 

\begin{equation}
{ \text{Let}  \hspace{.09cm}  G_{ab} := g_{ab} +  {\dfrac{2 \sigma}{m^2}} \hspace{.09cm}   R_{ab}-  {\dfrac{8 \zeta}{m^2}}  \hspace{.09cm}  R_{acbd}a^{c} a^{d} }
\end{equation}
Thus, the modified light cone condition can be written as 
\begin{equation}
  G_{ab}  k^{a} k^{b} = 0
\end{equation}
 Arguing along similar lines  as in the previous section, we get the dual polynomial and the principal polynomial for the system as
\begin{equation}
 {  P^{\#} (x,v) = G_{ab} v^{a} v^{b} }.
\end{equation}

\begin{equation}
 {  P (x,k) = G^{ab} k_{a} k_{b} }.
\end{equation}
 where $ G^{ab} $ is the inverse of the metric $ G_{ab} $.  Thus, we see that the correct relationship between $ p^{\nu}$ and  $k_{\mu}$ is:
\begin{equation}
p^{a} =  \dfrac{\partial ( G^{ab} k_{a} k_{b})}{\partial k_{a}} = G^{ab} k_{b}
\end{equation}
  Thus, $G^{ab}$ can be treated as an effective metric to the manifold.

Now, let us focus on the predictability of the modified Maxwell's equations. This condition immediately translates to the hyperbolicity of dual polynomial. Thus, there should exist some vector field $h$ such that $ P^{\#}(h)  >0$ and for any vector field q, the following equation has only two real roots
\begin{equation}
 {  P^{\#} (x,h + \lambda q) =0 }.
\end{equation}
 Expanding the equation in terms of $ G_{ab} $ results in a quadratic equation of $ \lambda $ as in Ref.\cite{fps14}. Thus, the condition boils down to the discriminant of the quadratic equation being positive i.e.
\begin{equation}
 { G_{ab} h^{a} q^{b} -  G_{ab} h^{a} h^{b }  \hspace{.09cm}G_{cd} q^{c} q^{d} >0 }.
\end{equation} 

  Let us choose a vector basis \{$\epsilon_{\alpha}$\} such that $\epsilon_{0} := h $ and $ G_{ab} \hspace{.09cm}  \epsilon^{a}_{0} \epsilon^{b }_{\alpha} = 0 $

As $ G_{ab}\epsilon^{a}_{0}\epsilon^{b }_{0} > 0$, the discriminant becomes  $ q^{\alpha} q^{\beta} \hspace{.09cm}  G_{ab}  \hspace{.09cm} \epsilon^{a}_{\alpha} \epsilon^{b }_{\beta}< 0 $.
As $ g_{ab}$ is lorentzian, we find that if $ g_{ab} h^{a} h^{b} >0$ , then $ q^{\alpha} q^{\beta} \hspace{.09cm}  g_{ab}  \hspace{.09cm} \epsilon^{a}_{\alpha} \epsilon^{b }_{\beta}< 0 $
\begin{equation}
 \Rightarrow {\dfrac{2 \sigma}{m^2}}  \hspace{.09cm}  R_{\alpha \beta} -  {\dfrac{8 \zeta}{m^2}} \hspace{.09cm}   R_{\alpha c \beta d} \hspace{.09cm} a^{c} a^{d} < 0 \label{eq:one}
\end{equation}

\begin{equation}
 {\dfrac{2 \sigma}{m^2}} \hspace{.09cm}   R_{00}-  {\dfrac{8 \zeta}{m^2}} \hspace{.09cm}   R_{0 c 0d} \hspace{.09cm} a^{c} a^{d} > 0 \label{eq:two}
\end{equation}
 Here $\alpha,  \beta$ runs from 1 to 3 and   $R_{\alpha \beta},  R_{\alpha c \beta d}$ are the components of the tensors w.r.t basis  \{$\epsilon_{\alpha}$\}

Multiplying inequality Eq.~(\ref{eq:one}) by  $g^{\alpha \beta}$ and multiplying inequality Eq.~(\ref{eq:two}) by  $g^{00}$  summing up the indices, we get,
\begin{equation}
{\dfrac{2 \sigma}{m^2}}  g^{\alpha \beta} \hspace{.09cm}  R_{\alpha \beta} -  {\dfrac{8 \zeta}{m^2}} g^{\alpha \beta} \hspace{.09cm}   R_{\alpha c \beta d} \hspace{.09cm} a^{c} a^{d} > 0
\end{equation}
\begin{equation}
  {\dfrac{2 \sigma}{m^2}} g^{00} \hspace{.09cm}   R_{00}-  {\dfrac{8 \zeta}{m^2}} g^{00} \hspace{.09cm}   R_{0 c 0d} \hspace{.09cm} a^{c} a^{d} > 0
\end{equation}
 These inequalities are satisfied if 
\begin{equation}
 \hspace{.09cm} {\dfrac{2 \sigma}{m^2}}  \hspace{.09cm}   R > {\dfrac{8 \zeta}{m^2}} \hspace{.09cm}   R_{ c d} \hspace{.09cm} a^{c} a^{d}  \end{equation}

Conversely, if inequality (3) is satisfied along with the condition that the metric  $g_{ab}$ is lorentzian, the matter equations will have  a well posed initial value problem.

\subsection{ The Lagrangian}
Using the trick that we employed in the previous section, we get the Lagrangian for the dynamics of the metric as :

\begin{equation}
 \therefore \mathcal{L}= \int  \left[ {\dfrac{1}{2\kappa}} \left( \mathcal{R} - 2 \Lambda \right) + \mathcal{L}_\mathrm{M} \right] \sqrt{-g}\, \mathrm{d}^4 x   \hspace{.09cm},  \text{where}  \hspace{.09cm}{\large{\kappa = 8 \pi G}}
\end{equation}
$\mathcal{R}$ is the Ricci tensor obtained from the metric $ G_{ab}$, $\Lambda$ is the cosmological constant and $ \mathcal{L}_\mathrm{M}$ is the Lagrangian of the matter fields appearing in the theory.

$\mathcal{R}$ contains the Ricci scalar of the metric  $g_{ab}$ and higher order derivatives of Ricci and Riemann--Christoffel tensor of the  metric  $g_{ab}$ .

 The equation of motion of the system can be obtained by varying the Lagrangian w.r.t the metric $g_{ab}$. As the $\mathcal{R}$ term is messy, it is difficult to approach the problem directly. So, we vary the Lagrangian with  $G_{ab}$ and then change the variation from $G_{ab}$ to $g_{ab}$. 

Focusing on the gravitational part of the Lagrangian and varying w.r.t to $G_{ab}$, we get

\begin{equation}
{  { \hspace{.2cm} \int 
        \left[ 
           {\dfrac{1}{2\kappa} } \left( \dfrac{\delta R}{\delta G_{ab}} +
             \dfrac{R}{\sqrt{-G}} \dfrac{\delta \sqrt{-G}}{\delta G_{ab} } 
            \right)  
        \right] \delta G_{ab} \sqrt{-G}\, \mathrm{d}^4x}} \hspace{1.5cm} 
\end{equation}
However, from the definition of $G_{ab}$. we have 

\begin{equation}
  \delta G_{ab} =  \delta g_{ab} +  {\dfrac{2 \sigma}{m^2}} \hspace{.09cm}   \delta R_{ab}-  {\dfrac{8 \zeta}{m^2}}  \hspace{.09cm}   \delta R_{acbd}a^{c} a^{d}          
\end{equation}

We analyze each term separately.
We know that from Ref. \cite{wein72}
\begin{equation}
\delta R^h{}_{cld} =\nabla_l (\delta \Gamma^h_{dc}) - \nabla_d (\delta \Gamma^h_{lc})
\end{equation}
\begin{equation}
 \text{and}\hspace{.2cm} \delta {\Gamma^{\alpha}}_{\beta \mu}=  \dfrac 12 g^{\alpha \lambda} (\nabla_{\beta} \delta g_{\lambda \mu}+ \nabla_{\mu} \delta g_{\beta \lambda}-
\nabla_{\lambda} \delta g_{\beta \mu})  
\end{equation}
\begin{eqnarray}
\therefore  \hspace{.09cm} \delta R^h{}_{cld} =\nabla_l ( \dfrac 12 g^{he} (\nabla_{d} \delta g_{ec}+ \nabla_{c} \delta g_{de}-
\nabla_{e} \delta g_{dc}) ) \nonumber \\
 - \nabla_d ( \dfrac 12 g^{he} (\nabla_{l} \delta g_{ec}+ \nabla_{c} \delta g_{le}-
\nabla_{e} \delta g_{lc}) )~.
\end{eqnarray}
\begin{eqnarray}
 \hspace{1.6cm} = \dfrac 12 g^{he} (\nabla_l \nabla_{d} \delta g_{ec}+\nabla_l  \nabla_{c} \delta g_{de}-
\nabla_l \nabla_{e} \delta g_{dc} \nonumber \\
 - \nabla_d \nabla_{l} \delta g_{ec}+ \nabla_d \nabla_{c} \delta g_{le}-
\nabla_d  \nabla_{e} \delta g_{lc}  )~ ~ .
\end{eqnarray}
\begin{equation}
\text{ And} \hspace{0.09cm}   \delta R_{acbd} = \delta (g_{kh}   R^h{}_{cld}) =  \delta g_{kh}   R^h{}_{cld}+   g_{kh}  \delta  R^h{}_{cld}  , 
\end{equation}

From pallatani identities [8], we have
\begin{eqnarray}
\delta R_{kl} =\dfrac 12 g^{ab} (\nabla_a \nabla_{l} \delta g_{kb} +\nabla_a \nabla_{k} \delta g_{lb} \nonumber \\
-\nabla_k \nabla_{l} \delta g_{ab} - \nabla_a \nabla_{b} \delta g_{kl}  ) ~ .
\end{eqnarray}

Therefore the variation can be written as
\begin{eqnarray}
\delta \mathcal{L} = { \int  {1 \over 2\kappa}( \mathcal{G}^{kl} \hspace{.09cm} \delta g_{kl} +  \mathcal{G}^{kl} \hspace{.09cm} {\dfrac{2 \sigma}{m^2}} \hspace{.09cm}   \delta R_{kl}} \nonumber \\
 -  \mathcal{G}^{kl} \hspace{.09cm} {\dfrac{8 \zeta}{m^2}}  \hspace{.09cm}   \delta R_{kcld}a^{c} a^{d}) \hspace{.09cm} \sqrt{-G}\,\hspace{.09cm}\mathrm{d}^4x  ~ .
\end{eqnarray} 

 where $ \mathcal{G}^{kl}$ is the Einstein tensor associated with metric $G_{ab}$. Consider the second term of the integral.

 $ { \int {1 \over 2\kappa}\mathcal{G}^{kl}{\dfrac{2 \sigma}{m^2}} \hspace{.09cm}   \delta R_{kl}  \sqrt{-G}\,\hspace{.09cm} \mathrm{d}^4x} $ 

\begin{eqnarray}
= \int {1 \over 2\kappa}\mathcal{G}^{kl} \hspace{.09cm} {\dfrac{2 \sigma}{m^2}} \hspace{.09cm} ( \dfrac 12 g^{ab} (\nabla_a \nabla_{l} \delta g_{kb}+\nabla_a \nabla_{k} \delta g_{lb} \nonumber \\
 -\nabla_k \nabla_{l} \delta g_{ab} - \nabla_a \nabla_{b} \delta g_{kl}  ) ) \hspace{.09cm} \sqrt{-G}\,\hspace{.09cm} \mathrm{d}^4x  ~ .
\end{eqnarray} 

Now, we use integration by parts twice to shift the covariant derivative from the variation of the metric and modify the dummy indices to get :
\begin{flushleft}
\hspace{1cm} $ { \int {1 \over 2\kappa} {\dfrac{2 \sigma}{m^2}} \hspace{.09cm} \nabla_b \nabla_{a}( g^{al} \hspace{.09cm} \mathcal{G}^{bk} - \dfrac 12 g^{kl} \hspace{.09cm} \mathcal{G}^{ab}- \dfrac 12 g^{ab} \hspace{.09cm} \mathcal{G}^{kl}) \hspace{.09cm} \delta g_{kl} \hspace{.09cm} \sqrt{-G}\,\hspace{.09cm} \mathrm{d}^4x}$ .
\end{flushleft}
 Now consider the third term
\begin{flushleft}
$ { \int {1 \over 2\kappa}\mathcal{G}^{kl} \hspace{.09cm} {\dfrac{8 \zeta}{m^2}}  \hspace{.09cm}   \delta R_{kcld}a^{c} a^{d}  \sqrt{-G}\,\hspace{.09cm} \mathrm{d}^4x} $ 
$= { \int {1 \over 2\kappa}\mathcal{G}^{kl} \hspace{.09cm} {\dfrac{8 \zeta}{m^2}}  \hspace{.09cm}  (  \delta g_{kh}   R^h{}_{cld}+   g_{kh}  \delta  R^h{}_{cld} )a^{c} a^{d}  \sqrt{-G}\,\hspace{.09cm} \mathrm{d}^4x} $
\end{flushleft}

Let us focus on the second term of the above expression i.e.

$ { \int {1 \over 2\kappa}\mathcal{G}^{kl} \hspace{.09cm} {\dfrac{8 \zeta}{m^2}}  \hspace{.09cm}  g_{kh}  \delta  R^h{}_{cld} a^{c} a^{d}  \sqrt{-G}\,\hspace{.09cm} \mathrm{d}^4x} \nonumber $

\begin{eqnarray}
=  \int {1 \over 2\kappa}\mathcal{G}^{kl} \hspace{.09cm} {\dfrac{8 \zeta}{m^2}} \hspace{.09cm} g_{kh} ( \dfrac 12 g^{he} (\nabla_l \nabla_{d} \delta g_{ec}+\nabla_l \nabla_{c} \delta g_{de}\nonumber \\
 -\nabla_l  \nabla_{e}  \delta g_{dc} -  \nabla_d \nabla_{l} \delta g_{ec}\nonumber \\
+ \nabla_d \nabla_{c} \delta g_{le} -\nabla_d \nabla_{e} \delta g_{lc} )) a^{c} a^{d} \sqrt{-G}\,\hspace{.09cm} \mathrm{d}^4x . ~ ~ ~ 
\end{eqnarray} 

Using the identity $\delta g_{ec}{}_{;ld}-\delta g_{ec}{}_{;dl}=R^a{}_{eld}\delta g_{ac} + R^a{}_{cld}\delta g_{ae}$ for first and fourth terms and using integration by parts twice on other terms and then changing the dummy indices, we get

${ \int {1 \over 2\kappa}\mathcal{G}^{kl} \hspace{.09cm} {\dfrac{8 \zeta}{m^2}}  \hspace{.09cm}  g_{kh}  \delta  R^h{}_{cld} a^{c} a^{d}  \sqrt{-G}\,\hspace{.09cm} \mathrm{d}^4x} $ 
\begin{eqnarray}
=  \int {1 \over 2\kappa}(\mathcal{G}^{ac} \hspace{.09cm} {\dfrac{4 \zeta}{m^2}}  \hspace{.09cm}  R^k{}_{acd} a^{l} a^{d} + \mathcal{G}^{la} \hspace{.09cm} {\dfrac{4 \zeta}{m^2}}  \hspace{.09cm}  R^k{}_{cad} a^{c} a^{d}\nonumber \\
 - \nabla_d \nabla_{c}( \mathcal{G}^{dc} \hspace{.09cm} {\dfrac{4 \zeta}{m^2}}  \hspace{.09cm}a^{l} a^{k})+\nabla_c \nabla_{d}( \mathcal{G}^{ld} \hspace{.09cm} {\dfrac{4 \zeta}{m^2}  \hspace{.09cm}a^{c} a^{k}) ) }\nonumber \\
   - \nabla_c \nabla_{d}( \mathcal{G}^{ck} \hspace{.09cm} {\dfrac{4 \zeta}{m^2}}  \hspace{.09cm}a^{l} a^{d})+\nabla_c \nabla_{d}( \mathcal{G}^{kl} \hspace{.09cm} {\dfrac{4 \zeta}{m^2}}  \hspace{.09cm}a^{c} a^{d}) ) \nonumber \\
  {\delta g_{kl} \sqrt{-G}\,\hspace{.09cm} \mathrm{d}^4x}) . ~ ~ 
\end{eqnarray} 

\begin{widetext}

$ \therefore$ The total variation is
\begin{center}
   $  \delta \mathcal{L} = { \int {1 \over 2\kappa}( \mathcal{G}^{kl} \hspace{.09cm} + {\dfrac{2 \sigma}{m^2}} \hspace{.09cm} \nabla_b \nabla_{a}( g^{al} \hspace{.09cm} \mathcal{G}^{bk}}- \dfrac 12 g^{kl} \hspace{.09cm} \mathcal{G}^{ab}- \dfrac 12 g^{ab} \hspace{.09cm} \mathcal{G}^{kl}) \hspace{.09cm} $
\end{center}
 \begin{center}
$ - \mathcal{G}^{ac} \hspace{.09cm}{\dfrac{4 \zeta}{m^2}} \hspace{.09cm}  R^k{}_{acd} a^{l} a^{d}  - \mathcal{G}^{la} \hspace{.09cm} {\dfrac{4 \zeta}{m^2}}  \hspace{.09cm}  R^k{}_{cad} a^{c} a^{d}  - \mathcal{G}^{kh} \hspace{.09cm} {\dfrac{8 \zeta}{m^2}}  \hspace{.09cm}  R^l{}_{chd} a^{c} a^{d}$
\end{center}
\begin{center}
 $- \nabla_d \nabla_{c}( \mathcal{G}^{dc} \hspace{.09cm} {\dfrac{4 \zeta}{m^2}}  \hspace{.09cm}a^{l} a^{k}) +\nabla_c \nabla_{d}( \mathcal{G}^{ld} \hspace{.09cm} {\dfrac{4 \zeta}{m^2}}  \hspace{.09cm}a^{c} a^{k} )$
\end{center}
\begin{center}
${+ \nabla_c \nabla_{d}( \mathcal{G}^{ck} \hspace{.09cm} {\dfrac{4 \zeta}{m^2}}  \hspace{.09cm}a^{l} a^{d}) -\nabla_c \nabla_{d}( \mathcal{G}^{kl} \hspace{.09cm} {\dfrac{4 \zeta}{m^2}}  \hspace{.09cm}a^{c} a^{d}) ) {\delta g_{kl} \sqrt{-G}\,\hspace{.09cm} \mathrm{d}^4x} } $ .
\end{center}
\begin{equation}
\stackrel{!}{=} 0 \hspace{.09cm} \text{( By principle of least action.)}
\end{equation}
 Thus, the integrand must vanish identically at all points. Therefore, the modified Einstein's equation is 

\begin{equation}
\mathcal{G}^{kl} \hspace{.09cm} + {\dfrac{2 \sigma}{m^2}} \hspace{.09cm} \nabla_b \nabla_{a}( g^{al} \hspace{.09cm} \mathcal{G}^{bk} -  \dfrac 12 g^{kl} \hspace{.09cm} \mathcal{G}^{ab}- \dfrac 12 g^{ab} \hspace{.09cm} \mathcal{G}^{kl})  - \mathcal{G}^{ac} \hspace{.09cm}{\dfrac{4 \zeta}{m^2}}  \hspace{.09cm}  R^k{}_{acd} a^{l} a^{d} - \mathcal{G}^{la} \hspace{.09cm} {\dfrac{4  \zeta}{m^2}}  \hspace{.09cm}  R^k{}_{cad} a^{c} a^{d} \nonumber
\end{equation}

\begin{equation}
 { - \mathcal{G}^{kh} \hspace{.09cm} {\dfrac{8 \zeta}{m^2}}  \hspace{.09cm}  R^l{}_{chd} a^{c} a^{d}  - \nabla_d \nabla_{c}( \mathcal{G}^{dc} \hspace{.09cm} {\dfrac{4 \zeta}{m^2}}  \hspace{.09cm}a^{l} a^{k}) +\nabla_c \nabla_{d}( \mathcal{G}^{ld} \hspace{.09cm} {\dfrac{4 \zeta}{m^2}}  \hspace{.09cm}a^{c} a^{k} )} \nonumber
\end{equation}

\begin{equation}
 {+ \nabla_c \nabla_{d}( \mathcal{G}^{ck} \hspace{.09cm} {\dfrac{4 \zeta}{m^2}}  \hspace{.09cm}a^{l} a^{d}) -\nabla_c \nabla_{d}( \mathcal{G}^{kl} \hspace{.09cm} {\dfrac{4 \zeta}{m^2}}  \hspace{.09cm}a^{c} a^{d})  = \kappa T^{kl}  }
\end{equation}
Where the $  T^ {kl}$ is the stress-energy tensor.
\end{widetext}
\subsection{Corrections to Schwarzschild metric }
  Solving directly for a homogeneous, static and spherically symmetric spacetime turns out to be a very difficult problem.Therefore, we linearize the metric and obtain the solutions to the modified equations. Thus, the metric is assumed to take the form 
\begin{equation}
 g = B(r) dt\otimes  dt+  A(r) dr\otimes  dr +  r^2( d\theta \otimes  d\theta +sin^2\theta  \hspace{.09cm} d\phi\otimes  d\phi )
\end{equation}
where $B(r) = -(1 +k (r)) $ and $ A(r) = 1 +j (r)$. Here  $k (r)$ and $j (r)$ are assumed to be infinitesimal in magnitude and we neglect their higher powers . We assume this applies to the higher order derivatives of the functions as well. In this way, we neglect the product or products of the function with their higher derivatives.This implies that the inverses are approximately given by
\begin{equation}
B ^{-1} (r) = -1 + k (r) \hspace{.09cm} \text{and}\hspace{.09cm}  A ^{-1} (r) = 1 -j (r)
\end{equation}
 However, upon close inspection of the modified Einstein's equations, we find that these differential equations are of the order of six in terms of  $k(r)$ and $j(r)$.  But we can bring the order of the equations down by using the fact that the we keep only terms that are linear in $k(r)$ and $j(r)$  and their derivatives.

 The calculation of the components of Riemann-Christoeffel tensor shows that all the non-vanishing terms are of infinitesimal magnitude. Hence, we can neglect the terms like  ${ \mathcal{G}^{kh} \hspace{.09cm} R^l{}_{chd} a^{c} a^{d}}$ as $ \mathcal{G}^{kh}  $ is also an infinitesimal quantity.
 Thus, the final equations comprise of terms containing $ \mathcal{G}^{kh}$  and $ a^{d}$ and their covariant derivatives.
 This motivates to us to start from $ G_{ab}$ rather than $ g_{ab}$. So , we define 
\begin{equation}
 G = \mathcal{B}(r) dt\otimes  dt+ \mathcal{A}(r) dr\otimes  dr \nonumber
\end{equation}
\begin{equation}
 \hspace{0.95cm} +  \mathcal{R}(r)^2( d\theta \otimes  d\theta +sin^2\theta  \hspace{.09cm} d\phi\otimes  d\phi )
\end{equation}

The component functions $\mathcal{B}(r)$, $\mathcal{A}(r)$ and $\mathcal{R}(r)$ can be calculated from the definition of $ G_{ab}$ in terms of $k (r)$ and $j (r)$, but instead we define it as the following 
\begin{equation}
\mathcal{B}(r) = 1 + v (r)  , \hspace{.09cm}  \mathcal{A} (r) = 1 +w (r) \label{metric}
\end{equation}
\begin{equation}
 \mathcal{R} (r) =r^2 +h (r)  
\end{equation}

and solve for $v (r)$, $w (r)$ and $h (r)$.

In the case of Schwarzschild solution for Einstein's field equations, we have the vanishing of all the components of the Einstein tensors as $  T^ {kl}$ is zero. Using this information and by inspecting our modified equations, we can make an educated guess for a solution given by 
\begin{equation}
  \mathcal{G}^{rr} =  \hspace{.09cm}\hspace{.09cm}\mathcal{G}^{\theta \theta } = 0
\end{equation}
 It follows from the symmetry of the metric that 
\begin{equation}
  \mathcal{G}^{\phi\phi} =  r^2  \hspace{.09cm} sin^2\theta  \hspace{.09cm} \mathcal{G}^{\theta \theta } = 0
\end{equation}
Thus the modified equations reduce to a single equation given by

\begin{equation}
\mathcal{G}^{tt} \hspace{.09cm}  + {\dfrac{\sigma}{m^2}} \nabla_t \nabla_{t}(  g^{tt} \hspace{.09cm} \mathcal{G}^{tt} )-  {\dfrac{2 \sigma}{m^2}}\nabla_b \nabla_{a}(\dfrac 12 g^{ab} \hspace{.09cm} \mathcal{G}^{tt})
\end{equation}
\begin{equation}
- \nabla_t \nabla_{t}( \mathcal{G}^{tt} \hspace{.09cm} {\dfrac{4 \zeta}{m^2}}  \hspace{.09cm}(a^{t})^2) +\nabla_c \nabla_{t}( \mathcal{G}^{tt} \hspace{.09cm} {\dfrac{4 \zeta}{m^2}}  \hspace{.09cm}a^{c} a^{t} ) \nonumber
\end{equation}

\begin{equation}
 {+ \nabla_t \nabla_{d}( \mathcal{G}^{tt} \hspace{.09cm} {\dfrac{4 \zeta}{m^2}}  \hspace{.09cm}a^{t} a^{d}) -\nabla_c \nabla_{d}( \mathcal{G}^{tt} \hspace{.09cm} {\dfrac{4 \zeta}{m^2}}  \hspace{.09cm}a^{c} a^{d})  = 0 }
\end{equation}

Thus, it is clear that the solution to the modified Einstein's equation depends upon the polarization of the photon. To obtain a nice analytic solution and for the sake of simplicity, we look at a case where $ a^{r} =(i {m^2})/{4 \zeta}  $ and  $ a^{\theta } = a^{\phi } = 0 $. Thus, the equation becomes

\begin{eqnarray}
  \mathcal{G}^{tt} +  {\dfrac{\sigma}{m^2}} \nabla_t \nabla_{t}(  g^{tt} \hspace{.09cm} \mathcal{G}^{tt} )-  {\dfrac{2 \sigma}{m^2}}\nabla_b \nabla_{a}(\dfrac 12 g^{ab} \hspace{.09cm} \mathcal{G}^{tt}) \nonumber \\
+ \nabla_r \nabla_{r}( \mathcal{G}^{tt} \hspace{.09cm}  {\dfrac{m^2}{4 \zeta}}  )= 0
\end{eqnarray}

Calculating the covariant derivative w.r.t $ g_{ab}$, and neglecting product of infinitesimal terms, we obtain
\begin{equation}
  \mathcal{G}^{tt} +\bigl{[}\dfrac{m^2}{4 \zeta}- {\dfrac{\sigma}{m^2}}\bigl{]}\hspace{.09cm}\dfrac{d^2 \mathcal{G}^{tt}}{d r^2}  -  {\dfrac{2 \sigma}{m^2}}\nabla_\theta \nabla_{\theta}( \dfrac{ \mathcal{G}^{tt}}{r^2} ) = 0
\end{equation}
This can be further simplified to get :
\begin{equation}
 \mathcal{G}^{tt} +\bigl{[}\dfrac{m^2}{4 \zeta}- {\dfrac{\sigma}{m^2}}\bigl{]}\hspace{.09cm} \dfrac{d^2 \mathcal{G}^{tt}}{d r^2}  -  {\dfrac{2 \sigma}{m^2}}\bigl{[} \dfrac{1}{r} \dfrac{d\mathcal{G}^{tt}}{d r} - \dfrac{2}{r^3} \dfrac{d \mathcal{G}^{tt}}{d r} +\dfrac{6 {G}^{tt}}{ r^4}  \bigl{]}= 0 \label{eq:ten}
\end{equation}

However, these equations are too complicated to have a well-behaved analytic solution. But, the equation can be reduced to a simpler form if we make some sensible approximations. Plugging in the values of the $\zeta , \sigma \hspace{.09cm}\text{and} \hspace{.09cm} m$, we find that  
\begin{equation}
  \dfrac{2\sigma}{m^2} \sim  10^{-32}  \bigl{[}\dfrac{m^2}{4 \zeta}- {\dfrac{\sigma}{m^2}}\bigl{]} \hspace{.09cm} \sim 10^{-16} 
\end{equation}

Thus, for very large $r$, the contribution of the third term in square bracket of Eq.~(\ref{eq:ten}) is negligible and we get the following equation :
\begin{equation}
 \mathcal{G}^{tt} +\bigl{[}\dfrac{m^2}{4 \zeta}- {\dfrac{\sigma}{m^2}}\bigl{]}\hspace{.09cm} \dfrac{d^2 \mathcal{G}^{tt}}{d r^2} = 0 \label{soln}
\end{equation}

This is an ordinary differential equation in one variable which can be solved easily. A physical solution is 
\begin{equation}
\mathcal{G}^{tt} =  e^{- k r} ,  \hspace{.09cm} \text{where} \hspace{.09cm}   k  = \frac{1}{\sqrt{\bigl{[}\dfrac{-m^2}{4 \zeta}+ {\dfrac{\sigma}{m^2}}\bigl{]}} }
\end{equation}
 Here  $ k$ is a constant and as $\zeta$ is negative, $k$ is real and positive. It is imperative that we check the consistency of the solution obtained before proceeding any further. We can easily see that only at very large $r$,  $\mathcal{G}^{tt}$ turns out to be infinitesimal in magnitude and this is necessary for our effective metric $G_{ab}$ to  satisfy the  Eq.~(\ref{metric})

This condition renders the contribution of the third term of Eq.~(\ref{eq:ten}) negligible and we end up getting  Eq.~(\ref{soln})

 Starting from the effective metric  $ G_{ab}$, we can calculate the components of the Einstein tensors given by  
\begin{equation}
  \mathcal{G}^{tt} = \dfrac{-w'}{r} +\dfrac{h''}{r^2} -\dfrac{h'}{2r^3} -\dfrac{3 h }{2 r^4}- \dfrac{w}{r^2} 
\end{equation}
\begin{equation}
  \mathcal{G}^{rr} = \dfrac{-v'}{r} -\dfrac{h'}{2r^3} -\dfrac{3 h }{2 r^4}+ \dfrac{w}{r^2} 
\end{equation}

\begin{equation}
 \mathcal{G}^{ \theta \theta} =-\dfrac{v''}{2 r^2} -\dfrac{(v'-w')}{2r^3} +\dfrac{3 h }{2 r^6}+\dfrac{h'}{2r^5} 
\end{equation}
\begin{equation}
  \mathcal{G}^{\phi\phi} =  r^2  \hspace{.09cm} sin^2\theta  \hspace{.09cm} \mathcal{G}^{\theta \theta } 
\end{equation}

Since the last three equations vanish, we can freely set $h(r)=0$ as $w(r) = r v'(r)$ solves both the second and third equation. Thus, first equation implies
\begin{equation}
  \dfrac{-w'}{r} - \dfrac{w}{r^2} =  e^{- k r}
\end{equation}
 This can be easily solved to get the solution

\begin{equation}
  w(r) = \dfrac{f}{r} + \dfrac{e^{-k r} \hspace{.09cm}\left(k^2 \hspace{.09cm}r^2+2\hspace{.09cm} k \hspace{.09cm}r+2\right)}{k^3\hspace{.09cm} r}
\end{equation}
 The constant of integration can be fixed by taking the classical limit of $r \to  \infty$ and identifying the leading order term of the metric component as that of the Schwarzschild metric for a very large $ r $ ,viz. $(1 + \dfrac{2  \hspace{.09cm} G \hspace{.09cm}M}{r})$. Thus, we get
\begin{equation}
  w(r) = \dfrac{2  \hspace{.09cm} G \hspace{.09cm}M}{r} + \dfrac{e^{-k r} \hspace{.09cm}\left(k^2 \hspace{.09cm}r^2+2\hspace{.09cm} k \hspace{.09cm}r+2\right)}{k^3\hspace{.09cm} r}
\end{equation}
 From $w(r) = r v'(r)$ , we get 
\begin{equation}
 v(r) = - \dfrac{2  \hspace{.09cm} G \hspace{.09cm}M}{r}  -\dfrac{e^{-k r} \hspace{.09cm} (k \hspace{.09cm} r+2)}{k^3 \hspace{.09cm} r}
\end{equation}
 Thus, we observe that the photon ``sees" a modified Schwarzschild metric, with a very small quantum correction. However, there is a peculiar property to this solution which is absent in the Schwarzschild solution of general relativity. Let us solve for the radius at which the radial component of velocity of the photon vanishes. This radial distance corresponds to the event horizon of the black hole. Since the photon travels along null geodesics in geometric optics approximation, we have

\begin{equation}
 G_{ab} \hspace{.09cm} u^{a}\hspace{.09cm} u^{b} = \mathcal{B}(r) \hspace{.09cm} \dot t^2+ \mathcal{A}(r) \hspace{.09cm} \dot r^2 =0
\end{equation}

 Here, $ u^{a}$ is the tangent vector along the trajectory of the photon, parametrized by some $\tau$ and the dot is taken w.r.t $\tau$ . The expression translates to
\begin{equation}
  \dfrac{dr}{dt} = \sqrt{- \mathcal{B}(r)/\mathcal{A}(r)}
\end{equation}
\begin{equation}
 ~~ ~~~~~~~~~~~~  = \sqrt{(1+v(r)) (1-w(r))} \label{hor}
\end{equation}
 The LHS is the radial component of the co-ordinate velocity of the photon. The dependence of the velocity with $r$ can analyzed using Eq.~(\ref{hor}). Plugging in the values of the functions, we can easily see that the radial component vanishes not at the Schwarzschild radius, but at a slightly closer value of $r$. This implies that the event horizon depends upon the direction of polarization of the light.

Since the metric $G_{ab}$ is lorentzian, the predictability of the matter equations is guaranteed and the principle polynomial becomes bi-hyperbolic. Thus, the solution doesn't violate the initial assumptions of the theory.

Now let us consider a photon which is not polarized in any particular direction. Thus, the components $a^{k}$ will be varying in both negate and positive directions. As the frequency of this variation is greater than the reaction rate of the metric to the fluctuations of the electromagnetic field, we take a statistical average and find that $\langle  a^{r} \rangle  \hspace{.09cm}=  \hspace{.09cm}\langle a^{t} \rangle \hspace{.09cm} =  \hspace{.09cm}\langle  a^{\theta} \rangle \hspace{.09cm} = \langle  a^{\phi} \rangle = 0$. Thus, if we set  $ \mathcal{G}^{rr} =  \hspace{.09cm}\hspace{.09cm}\mathcal{G}^{\theta \theta } = 0$, the new equations reduce to the Einstein's equations. Hence, we get back the Schwarzschild solution.

  This is an interesting phenomenon as some photons ``see" the event horizon at a particular distance while other photons with different polarization see the horizon at another radial co-ordinate.

This is related to the fact that the modified Einstein's equation is dependent on the polarization of the photon. This is analogous to the dependence of wavelength of light in the rainbow gravity theory [4]. This solution could also contribute to the resolution of Horizon problem arguing along the lines of Ref.\cite{smolin04}.

Thus, this particular examples provides a solution with features that are all well-known and provides a consistency check for our formalism.

\section{Discussion }
  In this paper, we have seen how the imposition of Cauchy predictability of an effective field theory can give rise to gravity. It is imperative that we understand where this new field is coming from. The existence of a well posed Cauchy problem is the most important part of any classical theory, even though one might have to specify a large number of initial conditions to actually solve the equations. As we have seen in the previous sections, the emergence of corrections to the metric was solely do to the imposition of such a constraint.

 A century of physics has taught us that our universe is quantum mechanical. If the emergent classical system were to approximately describe the same reality that the quantum theory did, it is only natural to think that the quantities that appear only in the classical theory must have some counterpart(s) in the quantum theory.

 Now returning to our specific example, we can easily see that if the fermionic systems were not interacting with the photons, tracing out the former wouldn't have any effect on the latter. This would not result in any corrections at all. Thus, gravity is related to quantum interactions that existed between the fermions and the bosons. 

 This deep connection of gravity with quantum interactions has been already been conjectured by a number of people like Van Raamsdonk, Maldacena and Susskind. But most of these ideas were in the light of string theory. However, in our formalism, we see such a connection arises without considering any assumptions pertaining to string theory.

In fact, we can also make the following observation. Effective field theories are perfectly local if we contain ourselves to energies lower than the mass of the heaviest fermionic particle of the theory. However, if we were to probe higher energies, we would end up getting highly non-local effects as we have traced out a particle lighter than the energy scales we are probing. In order to get the same physics at lower energies, we replace the non-local interactions from virtual heavy particle exchange with a set of local interactions in these theories. Thus, it plausible to assume that the corrections to metric might also have its roots in the non-local interactions that would have otherwise emerged if we probed higher energies. This relation might enable us to draw a parallel between quantum correlations and classical notions, owing to the generality of the procedure in the paper.

\nocite{*}

\bibliography{apssamp}% Produces the bibliography via BibTeX.

%merlin.mbs apsrev4-1.bst 2010-07-25 4.21a (PWD, AO, DPC) hacked
%Control: key (0)
%Control: author (8) initials jnrlst
%Control: editor formatted (1) identically to author
%Control: production of article title (-1) disabled
%Control: page (0) single
%Control: year (1) truncated
%Control: production of eprint (0) enabled
\providecommand{\noopsort}[1]{}\providecommand{\singleletter}[1]{#1}%
\begin{thebibliography}{18}%
\makeatletter
\providecommand \@ifxundefined [1]{%
 \@ifx{#1\undefined}
}%
\providecommand \@ifnum [1]{%
 \ifnum #1\expandafter \@firstoftwo
 \else \expandafter \@secondoftwo
 \fi
}%
\providecommand \@ifx [1]{%
 \ifx #1\expandafter \@firstoftwo
 \else \expandafter \@secondoftwo
 \fi
}%
\providecommand \natexlab [1]{#1}%
\providecommand \enquote  [1]{``#1''}%
\providecommand \bibnamefont  [1]{#1}%
\providecommand \bibfnamefont [1]{#1}%
\providecommand \citenamefont [1]{#1}%
\providecommand \href@noop [0]{\@secondoftwo}%
\providecommand \href [0]{\begingroup \@sanitize@url \@href}%
\providecommand \@href[1]{\@@startlink{#1}\@@href}%
\providecommand \@@href[1]{\endgroup#1\@@endlink}%
\providecommand \@sanitize@url [0]{\catcode `\\12\catcode `\$12\catcode
  `\&12\catcode `\#12\catcode `\^12\catcode `\_12\catcode `\%12\relax}%
\providecommand \@@startlink[1]{}%
\providecommand \@@endlink[0]{}%
\providecommand \url  [0]{\begingroup\@sanitize@url \@url }%
\providecommand \@url [1]{\endgroup\@href {#1}{\urlprefix }}%
\providecommand \urlprefix  [0]{URL }%
\providecommand \Eprint [0]{\href }%
\providecommand \doibase [0]{http://dx.doi.org/}%
\providecommand \selectlanguage [0]{\@gobble}%
\providecommand \bibinfo  [0]{\@secondoftwo}%
\providecommand \bibfield  [0]{\@secondoftwo}%
\providecommand \translation [1]{[#1]}%
\providecommand \BibitemOpen [0]{}%
\providecommand \bibitemStop [0]{}%
\providecommand \bibitemNoStop [0]{.\EOS\space}%
\providecommand \EOS [0]{\spacefactor3000\relax}%
\providecommand \BibitemShut  [1]{\csname bibitem#1\endcsname}%
\let\auto@bib@innerbib\@empty
%</preamble>
\bibitem [{\citenamefont {Maldacena}\ and\ \citenamefont
  {Susskind}(2013)}]{PROP:PROP201300020}%
  \BibitemOpen
  \bibfield  {author} {\bibinfo {author} {\bibfnamefont {J.}~\bibnamefont
  {Maldacena}}\ and\ \bibinfo {author} {\bibfnamefont {L.}~\bibnamefont
  {Susskind}},\ }\href {\doibase 10.1002/prop.201300020} {\bibfield  {journal}
  {\bibinfo  {journal} {Fortschritte der Physik}\ }\textbf {\bibinfo {volume}
  {61}},\ \bibinfo {pages} {781} (\bibinfo {year} {2013})}\BibitemShut
  {NoStop}%
\bibitem [{\citenamefont {Van~Raamsdonk}(2010)}]{VanRaamsdonk2010}%
  \BibitemOpen
  \bibfield  {author} {\bibinfo {author} {\bibfnamefont {M.}~\bibnamefont
  {Van~Raamsdonk}},\ }\href {\doibase 10.1007/s10714-010-1034-0} {\bibfield
  {journal} {\bibinfo  {journal} {Gen. Rel. Gravit.}\ }\textbf {\bibinfo
  {volume} {42}},\ \bibinfo {pages} {2323} (\bibinfo {year}
  {2010})}\BibitemShut {NoStop}%
\bibitem [{\citenamefont {Ryu}\ and\ \citenamefont
  {Takayanagi}(2006)}]{PhysRevLett.96.181602}%
  \BibitemOpen
  \bibfield  {author} {\bibinfo {author} {\bibfnamefont {S.}~\bibnamefont
  {Ryu}}\ and\ \bibinfo {author} {\bibfnamefont {T.}~\bibnamefont
  {Takayanagi}},\ }\href {\doibase 10.1103/PhysRevLett.96.181602} {\bibfield
  {journal} {\bibinfo  {journal} {Phys. Rev. Lett.}\ }\textbf {\bibinfo
  {volume} {96}},\ \bibinfo {pages} {181602} (\bibinfo {year}
  {2006})}\BibitemShut {NoStop}%
\bibitem [{\citenamefont {Maldacena}(1999)}]{Maldacena1999}%
  \BibitemOpen
  \bibfield  {author} {\bibinfo {author} {\bibfnamefont {J.}~\bibnamefont
  {Maldacena}},\ }\href {\doibase 10.1023/A:1026654312961} {\bibfield
  {journal} {\bibinfo  {journal} {Int. J. Theor. Phys.}\ }\textbf {\bibinfo
  {volume} {38}},\ \bibinfo {pages} {1113} (\bibinfo {year}
  {1999})}\BibitemShut {NoStop}%
\bibitem [{\citenamefont {Schuller}\ and\ \citenamefont {Witte}(2014)}]{fps14}%
  \BibitemOpen
  \bibfield  {author} {\bibinfo {author} {\bibfnamefont {F.~P.}\ \bibnamefont
  {Schuller}}\ and\ \bibinfo {author} {\bibfnamefont {C.}~\bibnamefont
  {Witte}},\ }\href {\doibase 10.1103/PhysRevD.89.104061} {\bibfield  {journal}
  {\bibinfo  {journal} {Phys. Rev. D}\ }\textbf {\bibinfo {volume} {89}},\
  \bibinfo {pages} {104061} (\bibinfo {year} {2014})}\BibitemShut {NoStop}%
\bibitem [{\citenamefont {R\"atzel}\ \emph {et~al.}(2011)\citenamefont
  {R\"atzel}, \citenamefont {Rivera},\ and\ \citenamefont {Schuller}}]{Rat10}%
  \BibitemOpen
  \bibfield  {author} {\bibinfo {author} {\bibfnamefont {D.}~\bibnamefont
  {R\"atzel}}, \bibinfo {author} {\bibfnamefont {S.}~\bibnamefont {Rivera}}, \
  and\ \bibinfo {author} {\bibfnamefont {F.~P.}\ \bibnamefont {Schuller}},\
  }\href {\doibase 10.1103/PhysRevD.83.044047} {\bibfield  {journal} {\bibinfo
  {journal} {Phys. Rev. D}\ }\textbf {\bibinfo {volume} {83}},\ \bibinfo
  {pages} {044047} (\bibinfo {year} {2011})}\BibitemShut {NoStop}%
\bibitem [{\citenamefont {Giesel}\ \emph {et~al.}(2012)\citenamefont {Giesel},
  \citenamefont {Schuller}, \citenamefont {Witte},\ and\ \citenamefont
  {Wohlfarth}}]{kris12}%
  \BibitemOpen
  \bibfield  {author} {\bibinfo {author} {\bibfnamefont {K.}~\bibnamefont
  {Giesel}}, \bibinfo {author} {\bibfnamefont {F.~P.}\ \bibnamefont
  {Schuller}}, \bibinfo {author} {\bibfnamefont {C.}~\bibnamefont {Witte}}, \
  and\ \bibinfo {author} {\bibfnamefont {M.~N.~R.}\ \bibnamefont {Wohlfarth}},\
  }\href {\doibase 10.1103/PhysRevD.85.104042} {\bibfield  {journal} {\bibinfo
  {journal} {Phys. Rev. D}\ }\textbf {\bibinfo {volume} {85}},\ \bibinfo
  {pages} {104042} (\bibinfo {year} {2012})}\BibitemShut {NoStop}%
\bibitem [{\citenamefont {Heisenberg}\ and\ \citenamefont
  {Euler}(2006)}]{physics/0605038}%
  \BibitemOpen
  \bibfield  {author} {\bibinfo {author} {\bibfnamefont {W.}~\bibnamefont
  {Heisenberg}}\ and\ \bibinfo {author} {\bibfnamefont {H.}~\bibnamefont
  {Euler}},\ }\href@noop {} {\  (\bibinfo {year} {2006})},\ \Eprint
  {http://arxiv.org/abs/arXiv:physics/0605038} {arXiv:physics/0605038}
  \BibitemShut {NoStop}%
\bibitem [{\citenamefont {Schneider}\ \emph {et~al.}(1992)\citenamefont
  {Schneider}, \citenamefont {Ehlers},\ and\ \citenamefont {Falco}}]{sch92}%
  \BibitemOpen
  \bibfield  {author} {\bibinfo {author} {\bibfnamefont {P.}~\bibnamefont
  {Schneider}}, \bibinfo {author} {\bibfnamefont {J.}~\bibnamefont {Ehlers}}, \
  and\ \bibinfo {author} {\bibfnamefont {E.}~\bibnamefont {Falco}},\
  }\href@noop {} {\emph {\bibinfo {title} {Gravitational Lenses}}}\ (\bibinfo
  {publisher} {Springer-Verlag},\ \bibinfo {year} {1992})\BibitemShut {NoStop}%
\bibitem [{\citenamefont {Shore}(2003)}]{shore03}%
  \BibitemOpen
  \bibfield  {author} {\bibinfo {author} {\bibfnamefont {G.~M.}\ \bibnamefont
  {Shore}},\ }\href {\doibase 10.1080/00107510310001617106} {\bibfield
  {journal} {\bibinfo  {journal} {Contemporary Physics}\ }\textbf {\bibinfo
  {volume} {44}},\ \bibinfo {pages} {503} (\bibinfo {year} {2003})},\ \Eprint
  {http://arxiv.org/abs/http://dx.doi.org/10.1080/00107510310001617106}
  {http://dx.doi.org/10.1080/00107510310001617106} \BibitemShut {NoStop}%
\bibitem [{\citenamefont {Drummond}\ and\ \citenamefont
  {Hathrell}(1980)}]{drum80}%
  \BibitemOpen
  \bibfield  {author} {\bibinfo {author} {\bibfnamefont {I.~T.}\ \bibnamefont
  {Drummond}}\ and\ \bibinfo {author} {\bibfnamefont {S.~J.}\ \bibnamefont
  {Hathrell}},\ }\href {\doibase 10.1103/PhysRevD.22.343} {\bibfield  {journal}
  {\bibinfo  {journal} {Phys. Rev. D}\ }\textbf {\bibinfo {volume} {22}},\
  \bibinfo {pages} {343} (\bibinfo {year} {1980})}\BibitemShut {NoStop}%
\bibitem [{\citenamefont {Weinberg}(1972)}]{wein72}%
  \BibitemOpen
  \bibfield  {author} {\bibinfo {author} {\bibfnamefont {S.}~\bibnamefont
  {Weinberg}},\ }\href@noop {} {\emph {\bibinfo {title} {Gravitation and
  cosmology: principles and applications of the general theory of
  relativity}}}\ (\bibinfo  {publisher} {Wiley},\ \bibinfo {year}
  {1972})\BibitemShut {NoStop}%
\bibitem [{\citenamefont {Magueijo}\ and\ \citenamefont
  {Smolin}(2004)}]{smolin04}%
  \BibitemOpen
  \bibfield  {author} {\bibinfo {author} {\bibfnamefont {J.}~\bibnamefont
  {Magueijo}}\ and\ \bibinfo {author} {\bibfnamefont {L.}~\bibnamefont
  {Smolin}},\ }\href {http://stacks.iop.org/0264-9381/21/i=7/a=001} {\bibfield
  {journal} {\bibinfo  {journal} {Classical and Quantum Gravity}\ }\textbf
  {\bibinfo {volume} {21}},\ \bibinfo {pages} {1725} (\bibinfo {year}
  {2004})}\BibitemShut {NoStop}%
\bibitem [{\citenamefont {Rovelli}(2004)}]{Car04}%
  \BibitemOpen
  \bibfield  {author} {\bibinfo {author} {\bibfnamefont {C.}~\bibnamefont
  {Rovelli}},\ }\href {\doibase 10.1017/CBO9780511755804} {\emph {\bibinfo
  {title} {Quantum Gravity}}},\ Cambridge Monographs on Mathematical Physics\
  (\bibinfo  {publisher} {Cambridge University Press},\ \bibinfo {year}
  {2004})\BibitemShut {NoStop}%
\bibitem [{\citenamefont {Barton}\ and\ \citenamefont
  {Scharnhorst}(1993)}]{Scharn93}%
  \BibitemOpen
  \bibfield  {author} {\bibinfo {author} {\bibfnamefont {G.}~\bibnamefont
  {Barton}}\ and\ \bibinfo {author} {\bibfnamefont {K.}~\bibnamefont
  {Scharnhorst}},\ }\href {http://stacks.iop.org/0305-4470/26/i=8/a=024}
  {\bibfield  {journal} {\bibinfo  {journal} {Journal of Physics A:
  Mathematical and General}\ }\textbf {\bibinfo {volume} {26}},\ \bibinfo
  {pages} {2037} (\bibinfo {year} {1993})}\BibitemShut {NoStop}%
\bibitem [{\citenamefont {Shore}(1996)}]{shore96}%
  \BibitemOpen
  \bibfield  {author} {\bibinfo {author} {\bibfnamefont {G.}~\bibnamefont
  {Shore}},\ }\href {\doibase http://dx.doi.org/10.1016/0550-3213(95)00646-X}
  {\bibfield  {journal} {\bibinfo  {journal} {Nuclear Physics B}\ }\textbf
  {\bibinfo {volume} {460}},\ \bibinfo {pages} {379 } (\bibinfo {year}
  {1996})}\BibitemShut {NoStop}%
\bibitem [{\citenamefont {Griffiths}(2012)}]{grif12}%
  \BibitemOpen
  \bibfield  {author} {\bibinfo {author} {\bibfnamefont {D.~J.}\ \bibnamefont
  {Griffiths}},\ }\enquote {\bibinfo {title} {Introduction to electrodynamics
  (4th edition)},}\ \ (\bibinfo  {publisher} {PHI Learning},\ \bibinfo {year}
  {2012})\ Chap.~\bibinfo {chapter} {4}, p.\ \bibinfo {pages} {197}\BibitemShut
  {NoStop}%
\bibitem [{\citenamefont {Hawking}(1976)}]{PhysRevD.14.2460}%
  \BibitemOpen
  \bibfield  {author} {\bibinfo {author} {\bibfnamefont {S.~W.}\ \bibnamefont
  {Hawking}},\ }\href {\doibase 10.1103/PhysRevD.14.2460} {\bibfield  {journal}
  {\bibinfo  {journal} {Phys. Rev. D}\ }\textbf {\bibinfo {volume} {14}},\
  \bibinfo {pages} {2460} (\bibinfo {year} {1976})}\BibitemShut {NoStop}%
\end{thebibliography}%

\end{document}